\documentclass[notoc,paper,letterpaper]{JHEP3} 
\usepackage{epsfig,amssymb}

\newcommand{\be}{\begin{equation}}
\newcommand{\ee}{\end{equation}}
\newcommand{\bea}{\begin{eqnarray}}
\newcommand{\eea}{\end{eqnarray}}
\newcommand{\ie}{{\it i.e.~}}
\newcommand{\eg}{{\it e.g.~}}

\def\I{\mathbb{I}}

\def\U{{\rm U}}
\def\SU{\mathop{\rm SU}}

\def\Sp{\mathop{\rm Sp}}
\def\tr{\mathop{\rm tr}}
\def\Pf{\mathop{\rm Pf}}
\def\t{\tilde}
\def\h{\hat}

\def\wh{\widehat}
\def\bar{\overline}
\def\del{{\partial}}

\def\a{\alpha}

\def\b{\beta}

\def\k{\kappa}
\def\g{\gamma}
\def\d{\delta}
\def\e{\epsilon}
\def\ve{\varepsilon}
\def\l{\lambda}
\def\L{\Lambda}
\def\s{\sigma}
\def\m{\mu}
\def\th{\theta}
\def\nf{{n_f}}
\def\nc{{n_c}}

\def\NN{{\cal N}}
\def\MM{{\cal M}}
\def\SS{{\cal S}}

\def\we{W_{\rm eff}}

\def\wep{{W^\ve_{\rm eff}}}

\def\Db{{\bar D}}
\def\dM{{\d M}}
\def\dMb{{\d\Mb}}
\def\Mb{{\bar M}}
\def\Mo{{M_0}}

\def\Mvo{{M^\v_0}}
\def\ad{{\dot\a}}
\def\mb{{\bar\m}}
\def\Mvo{{M^\ve_0}}
\def\dM{{\d M}}
\def\vb{{\bar\ve}}
\def\lb{{\bar\l}}
\def\thb{{\bar\th}}
\def\dS{{\d\Sigma}}

\title{Sp(N) higher-derivative F-terms via singular superpotentials}

\author{Philip C. Argyres and Mohammad Edalati\\
Physics Department, University of Cincinnati, Cincinnati OH 45221-0011\\
\email{argyres,edalati@physics.uc.edu}}

\abstract{We generalize the higher-derivative $F$-terms introduced
by Beasley and Witten \cite{bw0409} for $\SU(2)$ superQCD to 
$\Sp(\nc)$ gauge theories with fundamental matter.  We generate
these terms by integrating out massive modes at tree level from an 
effective superpotential on the chiral ring of the microscopic theory.
Though this superpotential is singular, its singularities are mild
enough to permit the unambiguous identification of its minima, and
gives sensible answers upon integrating out massive modes near any
given minimum.}

\begin{document}

\section{Introduction}

Supersymmetric gauge theories are more amenable to analysis than 
ordinary gauge theories (see \cite{is9509,p9702} for reviews). 
Until recently, much of the attention in supersymmetric gauge 
theories has been devoted to those with small numbers of flavors 
$\nf$. This is because for small $\nf$ exact results, such as effective 
superpotentials, can easily be guessed and verified using some 
relatively simple consistency checks \cite{s9309,s9402}.  This way of 
approaching the problem cannot easily be extended to larger numbers 
of flavors. Instead, one can reverse the strategy and start with 
the IR free regime of these theories where there are many massless 
flavors and strong quantum effects are negligible.  The strong 
quantum effects found in the lower-flavor theories are obtained 
from higher-derivative $F$-terms of a special form in these IR 
free theories upon integrating out flavors \cite{bw0409}.

These higher-derivative terms were calculated in \cite{bw0409} for
$\SU(2)$ superQCD with $\nf\geq 2$ fundamental flavors using one-instanton
methods.  In \cite{ae0510} we computed these terms by integrating out
massive modes at tree-level from an effective superpotential.  These
superpotentials are more singular than those normally considered:  the 
potentials derived from them have cusp-like singularities at their
minima.  However, these singularities are mild enough that they
unambiguously define the moduli space of vacua, and can be dealt with
analytically by means of a simple regularization procedure.  The
intuitive picture \cite{ae0510} is summarized in figure 1.  In
\cite{ae0511} we also computed such superpotentials for $\SU(\nc)$ 
superQCD with $\nf=\nc+2$.  

\FIGURE{\epsfig{file=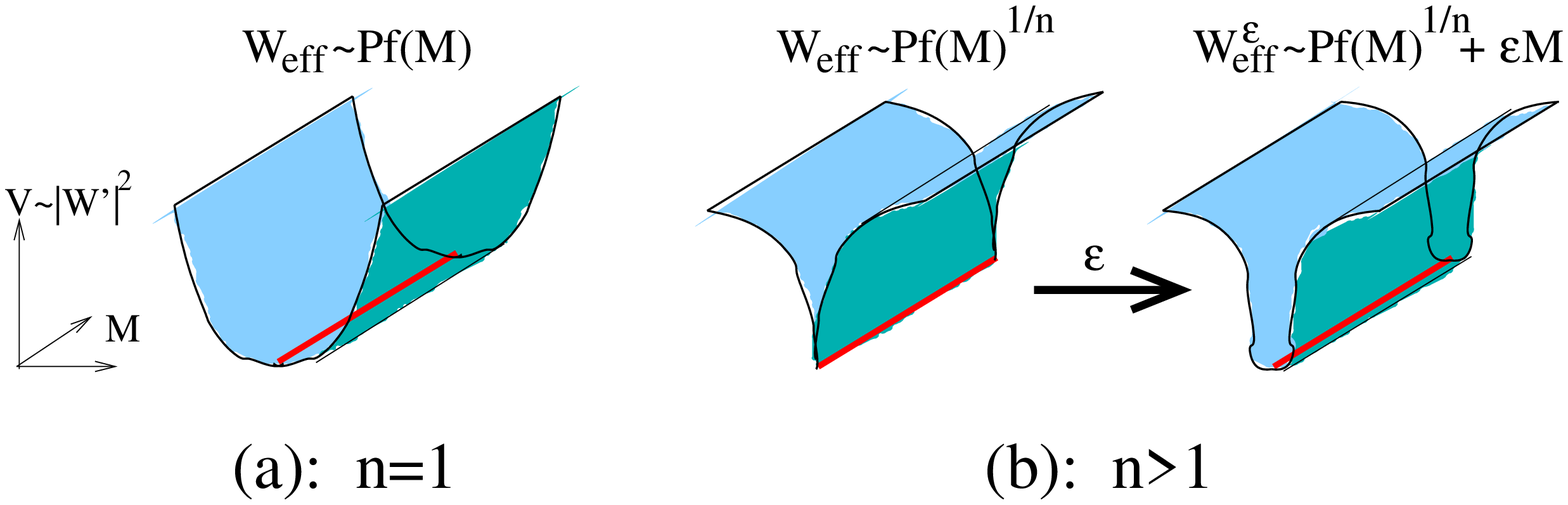,width=32em} 
\caption{The effective potential as a function of the meson
chiral fields $M$. The potential is regular for (a) $n=\nf-
\nc-1=1$, and singular for (b) $n>1$ where the cusp-like 
singularity can be smoothed by a regularization parameter $\ve$.}}

In this paper we generalize these results, and the results of 
\cite{bw0409}.  In section 2 we first compute the singular 
effective superpotentials of $\Sp(\nc)$ superQCD with 
matter in the fundamental representation, then we show that they correctly describe the moduli space of vacua, are consistent under RG flow 
to fewer flavors upon turning on masses, and are consistent solutions
to the Konishi anomaly equations \cite{k84,ks85}.  Then, in section
3, we generalize the results of \cite{bw0409} to $\Sp(\nc)$ 
superQCD by expanding the superpotential around a 
generic vacuum and integrating out the massive modes of the 
meson field  at tree level to find new higher-derivative $F$-terms.

\section{Large {$\bf \nf$} effective superpotentials of Sp($\bf\nc$) 
superQCD}

\subsection{Sp($\bf\nc$) superQCD for small $\bf\nf$}

Consider a four-dimensional $N=1$ $\Sp(\nc)$ supersymmetric gauge theory 
with $2\nf$ massless quark chiral fields $Q^i_a$ transforming in the 
fundamental representation, where $i=1, \ldots, 2\nf$ and $a=1, \ldots, 
2\nc$ are flavor and color indices, respectively.  (The number of flavors 
must be even for global anomaly cancellation \cite{w82}.)  The 
anomaly-free global symmetry of the theory is $\SU(2\nf) \times \U(1)_R$ 
under which the quarks transform as $({\bf 2\nf}, (\nf-\nc-1)/\nf)$. 

The classical moduli space of vacua is the space of vevs of holomorphic 
gauge-invariant chiral fields.  For $\Sp(\nc)$ superQCD these are the 
anti-symmetric meson fields $\h M^{[ij]}= Q^i_{a} J^{ab}Q^j_{b}$, where
$J^{ab}$ is the invariant antisymmetric tensor of $\Sp(\nc)$.  (We 
distinguish vevs from operators by hatting operators.)  For $\nf <\nc$
the classical moduli space is the space of arbitrary meson vevs $M^{ij}$, 
while for $\nf \ge \nc$ it is the set of all $M^{ij}$ subject to the 
condition rank$(M)\leq 2\nc$, or equivalently
\be\label{2.1}
\e_{i_1\cdots i_{2\nf}} M^{i_1 i_2} \cdots 
M^{i_{2\nc+1} i_{2\nc+2}} = 0.
\ee

Quantum mechanically there is a dynamically generated superpotential 
\cite{ip9505}
\be\label{2.2}
\we = \left\{
\begin{array}{ll}
(\nc+1-\nf) (\L^{b_0}/\Pf M )^{1/(\nc+1-\nf)},
&\qquad\mbox{for}\ 0<\nf\le\nc,\\
\Sigma\, (\Pf M - \L^{b_0}),
&\qquad\mbox{for}\ \nf=\nc+1,\\
-(\Pf M/\L^{b_0}), 
&\qquad\mbox{for}\ \nf=\nc+2,
\end{array}\right.
\ee
where the Pfaffian is defined as $\Pf M := \e_{i_1\cdots i_{2\nf}}
M^{i_1i_2}\cdots M^{i_{2\nf-1}i_{2\nf}}=\sqrt{\det M}$, $b_0 = 
3(\nc+1)-\nf$ is the coefficient of the one-loop beta function, 
$\L$ is the strong-coupling scale of the theory, and $\Sigma$ is
a Lagrange multiplier.  These superpotentials encode the low energy
behavior of the gauge theory: for $\nf\le\nc$ all the classical flat 
directions are lifted, for $\nf=\nc+1$ instantons deform the classical 
moduli space, while for $\nf=\nc+2$ the classical moduli space is not
modified.

\subsection{Superpotentials and classical constraints for large $\bf\nf$}

For $\nf>\nc+2$ the classical constraints are not modified, though 
there are new light degrees of freedom at singular subvarieties of
the moduli space when the theory is asymptotically free, $\nf<3\nc+3$.   
These singular subvarieties are commonly refered to as the ``origin" 
of the moduli space.  The only effective superpotential (for points away 
from the origin) consistent with holomorphicity, weak-coupling limits, 
and the global symmetries  is \cite{ip9505}
\be\label{spweff}
\we=-n\left(\Pf M\over \L^{b_0} \right)^{1/n},
\qquad
n:= \nf-\nc-1 >1.
\ee
The fractional power of $\Pf M$ implies that the potential derived from
this superpotential has cusp-like singularities at its extrema.  We will 
devote the rest of this section to arguing that, nevertheless, these 
singular superpotentials are physically perfectly sensible.

The first thing to check is to see whether these singular superpotentials 
describe the moduli space of vacua.   Because these superpotentials are 
singular at their extrema we cannot just naively extremize them.  We get 
around this problem by first deforming the superpotentials using some regularizing parameters $\ve_{ij}$, extremizing them, then taking the 
$\ve_{ij}\to 0$ limit at the end.   Independent of how the regularizing 
parameters are sent to zero, the extrema of the superpotentials must 
reproduce the classical constraint (\ref{2.1}).  The superpotentials
(\ref{spweff}) indeed pass this check, as we now show.

We regularize (\ref{spweff}) by adding a mass term with an invertible 
antisymmetric mass matrix $\ve_{ij}$ for the meson fields
\be\label{regweff}
\wep := \we + {1\over2}\ve_{ij}M^{ij}.
\ee
We have chosen to deform $\we$ by a linear term in $M^{ij}$ because it
is simple, it smooths the singularity, and it does not dominate at
large $M$, so does not introduce additional ``spurious" extrema.  We 
could have chosen a different deformation.  Varying $\wep$ with respect 
to $M^{kl}$ yields the equation of motion
\be\label{eom}
M^{kl}=-\L^{-b_0/n}(\Pf M)^{1/n}(\ve^{-1})^{kl}.  
\ee
Solving (\ref{eom}) for $\Pf M$ in terms of $\ve_{ij}$ and substituting back, 
we obtain 
\be\label{eomcont}
M^{kl}=-\L^{-{b_0/(\nc+1)}}(\Pf \ve)^{1/(\nc+1)}(\ve^{-1})^{kl}.
\ee
Multiplying $\nc+1$ copies of (\ref{eomcont}) together, and contracting 
the result with $\e_{i_1\ldots i_{2\nf}}$, we arrive at
\be\label{constdef}
\e_{i_1\ldots i_{2\nf}} M^{i_1 i_2} \cdots M^{i_{2\nc+1} i_{2\nc+2}} 
= {(-1)^{\nc+1}\over\L^{b_0}} \e_{i_1\ldots i_{2\nf}}
(\ve^{-1})^{i_1 i_2} \cdots (\ve^{-1})^{i_{2\nc+1} i_{2\nc+2}} \Pf \ve.
\ee
The right hand side of the above expression is a polynomial of order 
$n > 0$ in the $\ve_{ij}$. Therefore, in the  $\ve_{ij}\to 0$ limit it  
vanishes independently of how we take the limit and we have 
\be\label{consteq}
\e_{i_1\ldots i_{2N_f}} M^{i_1 i_2} \cdots M^{i_{2(N_c+1)-1} 
i_{2(N_c+1)}} = 0,
\ee
which is exactly the classical constraint that we wanted.  Furthermore, 
it is easy to check that all solutions of the classical constraints can 
be reached by taking $\ve_{ij}\to0$ appropriately.

Note that the negative power of $\L$ appearing in (\ref{spweff}) is 
not inconsistent with the weak coupling limit because the constraint 
equation (\ref{consteq}) which follows from extremizing the singular 
superpotential implies that $\Pf M=0$, thus $\we$ vanishes on the 
moduli space for any finite value of $\L$ as well as in the $\L\to0$ 
limit.  

We present another way of seeing how the classical constraints emerge 
from the singular superpotential which might make it clearer why
these superpotentials have unambiguous extrema.  Use the global 
symmetry to rotate the meson fields into the skew diagonal form  
\be\label{2.21}
M^{ij}=\pmatrix{M_1& & & \cr\ &M_2& & \cr\ & & \ddots & \cr\ &
& &M_{\nf}\cr}\otimes i\s_2,
\ee
so the effective superpotential (\ref{spweff}) becomes $\we = -n 
\L^{-b_0/n} (\prod_i M_i)^{1/n}$. The equations of motion which 
follow from extremizing with respect to the $M_i$ are
\be\label{2.23}
M_i^{{1\over n}-1}  \prod_{j\neq i} M_j^{1\over n}=0.
\ee
Though these equations are ill-defined if we set any of the $M_i=0$, 
we can probe the solutions by taking limits as some of the $M_i$ 
approach zero.  To test whether there is a limiting solution where 
$K$ of the $M_i$ vanish, consider the limit $\ve\to0$ with $M_1\sim
\ve^{\a_1}, \ldots, M_K\sim\ve^{\a_K}$ with $\a_j>0$ to be determined. 
Note that different non-zero values $\a_j$ corresponds to different 
deformations in (\ref{regweff}).  Substituting into (\ref{2.23}), 
only the first $K$ equations have non-trivial limits,
\be\label{2.24}
\lim_{\ve\to0}\ve^{{1\over n} \left(\sum_j \a_j\right)-\a_i}=0,
\qquad i=1,\ldots, K,
\ee
giving the system of inequalities $n\a_i< \sum_j \a_j$ for $i=1,\ldots,K$. 
These inequalities have solutions if and only if $K>n$, implying that 
rank$(M)\leq 2\nc$ which is precisely the classical constraint (\ref{2.1}).

\subsection{Consistency with Konishi anomaly equations: direct
description.}

In the previous section the effective superpotential 
of the theory was determined by the global symmetry, weak-coupling 
limits, and holomorphicity.  In this section we use the Konishi 
anomaly equations to derive the same superpotentials (\ref{spweff}). 
The Konishi anomaly \cite{k84,ks85} implies a set of differential 
equations which the effective superpotential should obey. 
We will show, using both the direct description and Seiberg dual 
description \cite{s9411,ip9505} of the theory, that the solution 
to the Konishi anomaly equations coincides with our 
singular superpotentials.  This is a consistency check on these 
superpotentials. 

It has been shown \cite{w0302} that for a pure superYang-Mills 
theory with the $\Sp(\nc)$ gauge group, the glueball superfield 
$\h S={1\over {32\pi^2}}\tr ({W^\a W_\a})$ generates all the 
local gauge-invariant chiral operators in the chiral ring 
of the theory.  When we add matter multiplets to a superYang-Mills 
theory we also need to include local gauge-invariant matter 
generators.  Following the arguments of \cite{cdsw0211,s0212} 
it follows that $\h S$ and $\h M^{ij}$ comprise all the local 
gauge-invariant chiral generators in the chiral ring.  In the 
chiral ring the Konishi anomaly for a tree level superpotential 
$W_{\rm tree}$ is
\be\label{kadir1}
\langle {\del W_{\rm tree}\over \del Q_a^i} Q_a^j\rangle = S \d^j_i.
\ee
The above set of equations are perturbatively one-loop exact and 
do not get non-perturbative corrections.  See \cite{s0311,ae0510,ae0511} 
for discussions on the non-perturbative exactness of the Konishi anomaly equations.  

As our tree level superpotential, we take  
\be\label{kadir2}
W_{\rm tree}= m_{ij}(\h M^{ij}-M^{ij}).
\ee
where $m_{ij}$ is a Lagrange multiplier enforcing $\h M^{ij}$ to have 
$M^{ij}$ as their vevs.  It follows from the form of the above 
tree-level superpotential and the nature of the Legendre transform 
\cite{is9509,ils9403,i9407} that
\be\label{kadir3}
m_{ij}=-{1\over 2}{{\del \we} \over  \del M^{ij}}.
\ee
Substituting (\ref{kadir2}) into (\ref{kadir1}) and using the fact 
that the expectation value of a product of gauge-invariant chiral 
operators equals the product of the expectation values, gives 
$2 m_{ik} M^{kj} = S \d^j_i$.  Using (\ref{kadir3}) we then obtain 
a set of partial differential equations for the effective superpotential
\be\label{kadir4}
{\del \we \over \del M^{ik}} M^{kj} = S \d^j_i,
\ee
whose solution is  
\be\label{kadir5}
\we(M, S) = S \ln \left(\Pf M \over \L^{\nf}\right)+ f(S), 
\ee
where the stong-coupling scale of the theory $\L$ has been 
inserted to make the quantity inside logarithm dimensionless.  The
function $f(S)$ is determined by the $\U(1)_R$ symmetry to be
\be\label{kadir6}
f(S) =  -nS\left[\ln (S/\L^3) - 1\right].
\ee
(The constant term in the brackets, which can be absorbed in a
re-definition of $\L$, was determined by matching to the
traditional normalization of the Veneziano-Yankielowicz 
superpotential \cite{vy82} after giving masses and integrating
out all the quarks.)

The glueball $S$ is massive away from the origin so can be 
integrated out.  Substituting (\ref{kadir6}) into (\ref{kadir5}) 
and integrating $S$ out by solving its equation of motion, we 
arrive at the effective superpotentials (\ref{spweff}).

\subsection{Consistency with Konishi anomaly equations: Seiberg 
dual description.}

In this subsection we use the Konishi anomaly approach for the 
dual description as well as the Seiberg duality dictionary to 
rederive once again our singular effective superpotentials.  For 
$\nf>\nc+2$ the theory has a Seiberg dual description as an 
$\Sp(\nf-\nc-2)$ supersymmetric gauge theory \cite{ip9505} with 
$2\nf$ (dual) quark chiral fields $q^{a}_i$, $i=1\ldots 2\nf$, in 
the fundamental representation and a gauge-singlet elementary 
field $\h\MM^{[ij]}$ coupled to the (dual) meson fields $\h \NN_{ij}
:=q^a_i J_{ab} q^b_j$ through the superpotential $W=\h\NN_{ij}
\h\MM^{ij}$.  The dual description is IR free when $\nf < {3\over2}
(\nc+2)$.

The ring of local gauge-invariant chiral operators for the dual 
theory is generated by the dual glueball superfield $\h\SS$, 
$\h\MM^{ij}$ and $\h\NN_{ij}$. 
As our tree level superpotential we take 
\be\label{kasd2}
W_{\rm tree} = \h\NN_{ij}\h\MM^{ij}+m_{ij}(\h\MM^{ij}-\MM^{ij}),
\ee
where $m_{ij}=-{1\over2}(\del\we/\del\MM^{ij})$ is the Lagrange 
multiplier associated with the dual description (not to be confused 
with Lagrange multiplier of the direct description).  It imposes the 
constraint that $\h \MM^{ij}$ have $\MM^{ij}$ as their vevs.  The 
superpotential $W=\h\NN_{ij}\h\MM^{ij}$ gives masses to the dual 
quarks and sets $\NN_{ij}=0$ when $\MM^{ij}\neq0$, which is why we 
have not included Lagrange multipliers for the dual mesons $\h\NN_{ij}$.

The Konishi anomaly equations for a tree level superpotential in 
the dual theory is  
$\langle (\del W_{\rm tree}/\del q^a_i) q^a_j\rangle = \SS \d^i_j$. 
Substituting (\ref{kasd2}) gives $2\MM^{ik}\NN_{kj}
=-\SS\d^i_j$.  Using the $\h\MM^{ij}$ equation of motion, $\NN_{ij}=
-m_{ij}$, we can eliminate $\NN_{kj}$ and arrive at 
\be\label{kasdcont}
\MM^{ik}{{\del\we}\over{\del\MM^{kj}}}=\SS\d^i_j, 
\ee
whose solution is
\be\label{kasd3}
\we(\MM, S) = \SS \ln \left({\Pf\MM \over \t\L^{\nf}}\right)+ g(\SS), 
\ee
where $\t\L$ is the strong-coupling scale of the dual theory.
$g(\SS)$ is determined as before to be $g(\SS) = -n\SS 
[\ln(\SS/\t\L^3)-1]$.  Integrating out $\SS$ then gives the effective 
superpotential in the dual description
\be\label{kasd4}
\we=n\left(\t\L^{3n-\nf}\Pf\MM \right)^{1/n}.
\ee 

The dual and the direct theories describe the same physics in the 
IR regime.  Both theories have the same global symmetries and, away 
from the origin, they have the same moduli space and the same light 
degrees of freedom.  They should also have the same effective 
superpotentials.  Thus, relabeling (\ref{kasd4}) in terms of the 
direct theory degrees of freedom, we should recover the singular 
superpotential of the direct theory.  In fact, using the Seiberg 
duality dictionary, the $\MM^{ij}$ are identified with the direct 
theory mesons through  $\MM^{ij}={1\over\m}M^{ij}$, where $\m$ is 
a mass scale related to the dual and the direct theory scales by
\be\label{2.14}
\L^{3(\nc+1)-\nf}\t\L^{3n-\nf} =(-1)^n \m^{\nf}.
\ee 
Using this, upon rewriting (\ref{kasd4}) in terms of $\L$ and $M^{ij}$ 
we indeed find the direct theory superpotential (\ref{spweff}).

\subsection{Consistency upon integrating out flavors.}

Besides correctly describing the moduli space, the effective 
superpotentials should also pass some other tests. If we add a 
mass term for one flavor in the superpotentials of a theory with 
$\nf$ flavors and then integrate it out, we should recover the 
superpotential of the theory with $\nf-1$ flavors.  To show that 
the effective superpotential (\ref{spweff}) passes this test, we 
add a gauge-invariant mass term for one flavor, say $M^{{2\nf-1}~{2\nf}}$,
\be\label{2.26}
\we =-n\L^{-b_0/n} (\Pf M)^{1/n}+ mM^{{2\nf-1}~{2\nf}}.
\ee  
The equations of motion for $M^{{i}~{2\nf-1}}$ and $M^{{j}~{2\nf}}$ 
($i\neq 2\nf-1$ and $j\neq 2\nf$) put the meson matrix into the form 
$M^{ij} = {\wh M \ 0 \choose 0 \ \wh X}$ where $\wh M$ is a $2(\nf-1)
\times 2(\nf-1)$ and $\wh X$ a $2\times2$ matrix.  Integrating out 
$\wh X\sim M^{{2\nf-1}~{2\nf}}$ by its equation of motion gives
\be\label{2.28}
\we =-(n-1)\wh\L^{b_0/(n-1)} (\Pf \wh M)^{1/(n-1)},
\ee
where $\wh\L = m \L^{3(\nc+1)-\nf}$ is the strong-coupling scale of 
the theory with $\nf-1$ flavors, consistent with matching the RG flow 
of couplings at the scale $m$.  Dropping the hats, we recognize 
(\ref{2.28}) as the effective superpotentials of $\Sp(\nc)$ superQCD 
with $\nf-1$ flavors.

\section{Higher-derivative F-terms in Sp($\bf\nc$) superQCD}

So far we have seen that our singular effective superpotentials 
(\ref{spweff}) correctly describe the moduli space of vacua.  In 
this section we will use these superpotentials to derive the form 
of certain higher-derivative $F$-terms in these theories.  This 
derivation can be taken as a prediction for the result of instanton 
calculations in the $\Sp(\nc)$ superQCD with large number of flavors.
 
In \cite{bw0409} Beasley and Witten showed that on the moduli space 
of $\SU(2)$ superQCD with $\nf\ge2$, instantons generate a series of 
higher-derivative $F$-terms (also called multi-fermion $F$-terms). 
As $F$-terms they are protected by non-renormalization theorems, and
so should be generated at tree level in perturbation theory from
an exact low energy effective superpotential.  Indeed, they 
also calculated these $F$-terms by integrating out massive modes 
at tree level from the non-singular effective superpotentials 
of $\SU(2)$ supersymmetic QCD with $\nf=2$ and $3$ flavors. 
In \cite{ae0510}, it was shown that singular effective 
superpotentials of $\SU(2)$ supersymmetic QCD can reproduce 
the corresponding $F$-terms for $\nf>3$, as well.  

We will show in this section that the singular superpotentials 
of $\Sp(\nc)$ superQCD (\ref{spweff}) likewise generate higher-derivative
$F$-terms by a tree-level calculation. As in our discussion of the 
classical constraints in the last section, the key point in this 
calculation is to first regularize the effective superpotentials 
(\ref{spweff}), and then show that the results are independent of 
the regularization.  

The $\SU(2)$ $F$-terms of \cite{bw0409} have the form
\bea\label{bw}
\d S &=& \int d^4x~d^2\th\,\L^{6-\nf}(M\Mb)^{-\nf}
\e^{i_1 j_1 \cdots i_\nf j_\nf}\Mb_{i_1 j_1} \nonumber\\
&&\quad\mbox{}\times
(M^{k_2 \ell_2} \Db\Mb_{i_2 k_2} \cdot \Db\Mb_{j_2 \ell_2}) \cdots 
(M^{k_\nf \ell_\nf} \Db\Mb_{i_\nf k_\nf} \cdot \Db\Mb_{j_\nf \ell_\nf}),
\eea
where $(M\Mb) := (1/2)\sum_{ij} M^{ij}\Mb_{ij}$, and the dot 
denotes contraction of the spinor indices on the covariant 
derivatives $\Db_\ad$. Although these terms are 
written in terms of the unconstrained meson field, they are to 
be understood as being evaluated on the classical moduli space.  
In other words, we should expand the $M^{ij}$ in (\ref{bw}) about 
a given point on the moduli space, satisfying $\e_{i_1\cdots 
i_{2\nf}} M^{i_1 i_2} \cdots M^{i_{2\nc+1} i_{2\nc+2}} = 0$, and 
keep only the massless modes (\ie those tangent to the moduli space).  
We will refer to such terms as being ``on vacuum" (in analogy to 
states being on mass-shell).  They should be contrasted with 
our effective superpotentials (\ref{spweff}) which are ``off 
vacuum".

Even though (\ref{bw}) is written as an integral over a 
chiral half of superspace, it is not obvious that the 
integrand is a chiral superfield.  But the form of the integrand 
is special: it is in fact chiral, and cannot be written as 
$\Db^2$(something), at least globally on the moduli space, and 
so is a protected term in the low energy effective action
\cite{bw0409}.  

\subsection{${\bf Sp(\nc)}$ F-terms}

To derive on-vacuum effective 
interactions from an off-vacuum term, we simply have to expand 
around a given point on the moduli space and integrate out the 
massive modes at tree level; see figure 2. The only technical 
complication is that the effective superpotential needs to be 
regularized first, \eg\ by turning on a small mass parameter 
$\ve_{ij}$ as in (\ref{regweff}), so that it is smooth at its 
extrema.  At the end, we take $\ve_{ij}\to 0$.  The absence of 
divergences as $\ve_{ij}\to0$ is another check of the consistency 
of our singular effective superpotentials.
What we will actually compute is just the leading $F$-term in an 
expansion around a generic point on the vacuum in terms of the massless 
modes of the meson (those tangent to the moduli space). 

\FIGURE{\epsfig{file=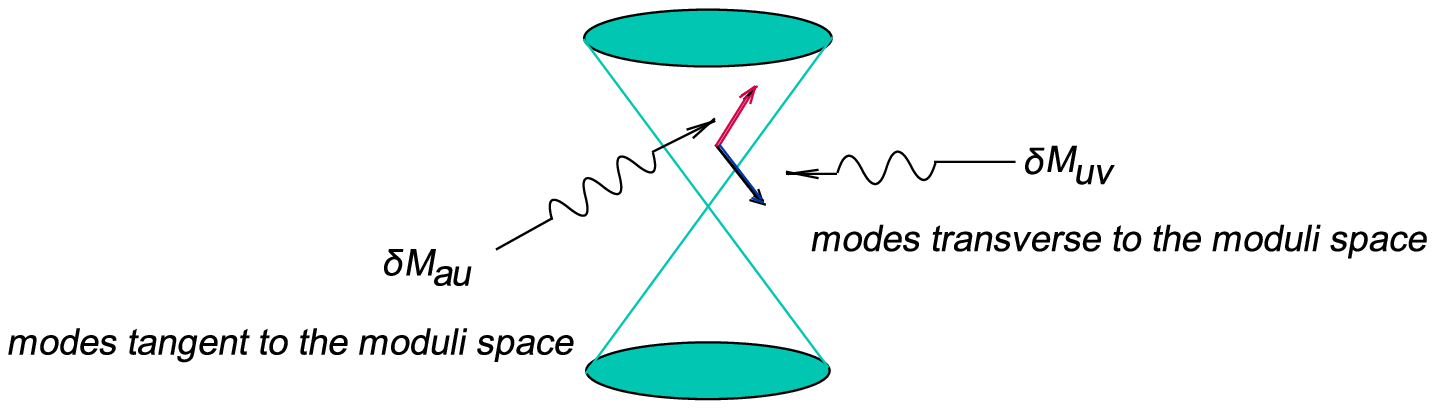,width=105mm}
\caption{The massless tangent modes $M_{au}$ (red arrow), 
and the massive transverse modes $M_{uv}$ (blue arrow) after 
the meson field $M_{ij}$ has been expanded around a given 
point on the moduli space.}}

As mentioned in section 2, the moduli space of vacua for $\Sp(\nc)$ 
superQCD with $\nf\ge\nc+2$ flavors is given by the constraint
\be\label{spnconstr}
\mbox{rank}(M^{ij}) \le 2\nc.
\ee
At a generic point on the moduli space, the vev of the meson field 
$\Mo^{ij} = \langle M^{ij}\rangle$ breaks the $\SU(2\nf)$ flavor 
symmetry. We can use flavor rotations to bring the generic vev 
into the form 
\be\label{spnvac}
\Mo^{ij}=\pmatrix{\m^{ab}&\cr &0\cr},
\ee
where $\m^{ab}=\mu \I_{\nc}\otimes i\sigma_2$ is a skew-diagonalized 
antisymmetric matrix and $\m$ is a complex parameter. Note that the 
above form for $\Mo^{ij}$ breaks the $\SU(2\nf)$ flavor symmetry to 
$\Sp(2\nc)\times\SU(2\nf-2\nc)$. Accordingly, we partition the 
$i,j,\ldots$ flavor indices into two sets: $\Sp(2\nc)$ indices 
$a,b,\ldots\in\{1,\ldots,2\nc\}$ from the front of the alphabet, 
and $\SU(2\nf-2\nc)$ indices $u,v,\ldots\in\{1,\ldots,2\nf-2\nc\}$ 
from the back.  Linearizing the meson field around (\ref{spnvac}), 
$M^{ij} = \Mo^{ij} + \dM^{ij}$, subject to the constraint 
(\ref{spnconstr}), implies that the massless modes are $\dM^{ab}$ 
and $\dM^{au}$, while the $\dM^{uv}$ are all massive, as in figure 2. 
The $\dM^{ab}$ modes can be absorbed in a change of $\m$, so we only 
need to focus on the $\dM^{au}$ modes. 

We will find that the 
leading $F$-term has the form
\bea\label{mfftlead}
\d S &\sim& \int d^4x~d^2\th\, \l^{-n} (\m\mb)^{-\nf}
\Pf(\mb)~\e^{u_1v_1\cdots u_{n+1}v_{n+1}}
\,(\m^{c_1 d_1}\Db\dMb_{c_1u_1} \cdot \Db\dMb_{d_1v_1})\times 
\nonumber\\ 
&&\cdots\times (\m^{c_{n+1} d_{n+1}}
\Db\dMb_{c_{n+1}u_{n+1}} \cdot \Db\dMb_{d_{n+1} v_{n+1}}),
\eea
where we have defined
\be
n:= \nf-\nc-1, \qquad\mbox{and}\qquad
\l := \L^{-b_0/n} = \L^{(\nf-3\nc-3)/n}.
\ee
Supersymmetry together with the flavor symmetry then uniquely 
determine the completion of this leading order term to all orders
to be
\bea\label{bwgen}
\d S &=& \int d^4x~d^2\th\,\L^{3\nc+3-\nf}(M\Mb)^{-\nf}
\e^{i_1 j_1 \cdots i_\nf j_\nf}
\Mb_{i_1 j_1} \cdots \Mb_{i_\nc j_\nc} \\
&&\quad\mbox{}\times
(M^{k_{\nc+1} \ell_{\nc+1}} \Db\Mb_{i_{\nc+1} k_{\nc+1}} 
\cdot \Db\Mb_{j_{\nc+1} \ell_{\nc+1}}) \cdots 
(M^{k_\nf \ell_\nf} \Db\Mb_{i_\nf k_\nf} \cdot \Db\Mb_{j_\nf \ell_\nf}).
\nonumber
\eea
This follows by an identical argument to one in \cite{bw0409}.
Indeed, (\ref{bwgen}) is a straightforward generalization of 
(\ref{bw}) and has many similar properties, including that it
is an $F$-term globally on the moduli space.

To generate the leading term (\ref{mfftlead}), we first regularize 
$\we \to \wep = -n\l (\Pf M)^{1/n} + {1\over2}\ve_{ij} M^{ij}$, and 
choose $\ve_{ij} = \l \ve^{1/n} \m^{2/n} \mbox{diag} \{\ve,1,\ldots,1\} 
\otimes i\s_2$ so that
\be\label{deformedvac}
(\Mvo)^{ij}=\pmatrix{\m^{ab}&\cr &\ve^{uv}\cr},
\ee
where $\ve^{uv}=\ve\I_{\nf-\nc}\otimes i\sigma_2$ is a $2(\nf-\nc)
\times 2(\nf-\nc)$ skew-diagonalized matrix. An advantage of this 
choice is that it preserves an $\Sp(2\nc) \times \Sp(2\nf-2\nc)$ 
subgroup of the flavor symmetry.  In the $\ve\to0$ limi, this 
symmetry is enhanced to $\Sp(2\nc)\times\SU(2\nf-2\nc)$.  Also, 
the massless directions around this choice of $(\Mvo)^{ij}$ are 
still $\dM^{ua}$ as before.

\subsection{Feynman rules}

We use standard superspace Feynman rules \cite{ggrs0108} to compute 
the effective action for the massless $\dM^{ua}$ modes by integrating 
out the massive $\dM^{uv}$ modes. This means we need to evaluate 
connected tree diagrams at zero momentum with internal massive 
propagators and external massless legs. In order to evalute these 
diagrams for the theory under discussion, we closely follow 
\cite{ae0510} where the superspace Feynman rules for $\SU(2)$ 
superQCD have been explained in detail.  Generalizing these 
rules for $\Sp(\nc)$ superQCD is easy:  the massive modes 
will have standard chiral, anti-chiral, and mixed superspace 
propagators with masses derived from the quadratic terms in the 
expansion of $\wep$, while higher-order terms in the expansion 
give chiral and anti-chiral vertices.  

A quadratic term in the superpotential, $W={1\over2}m(\dM)^2$, gives 
a mass which enters the chiral propagator as $\langle\dM\dM\rangle = 
\bar m (p^2+|m|^2)^{-1} (D^2/p^2)$, similarly for the anti-chiral 
propagator, and as $\langle\dM\dMb\rangle = (p^2+|m|^2)^{-1}$ 
for the mixed propagator.  Each propagator comes with a factor of 
$\d^4(\th -\th')$.  Even though the diagrams will be evaluated at 
zero momentum, we must keep the $p^2$-dependence in the above propagators 
for two reasons. First, there are spurious poles at $p^2=0$ in 
the (anti-)chiral propagators which will always cancel against 
momentum dependence in the numerator coming from $\Db^2$'s in the 
propagators and $D^2$'s in the vertices. For instance, $D^2\Db^2=p^2$ 
when acting on an anti-chiral field, giving a factor of $p^2$ in 
the numerator which can cancel that in the denominator of the 
anti-chiral propagator, to give an IR-finite answer.  Second, expanding 
the IR-finite parts in a power series in $p^2$ around $p^2=0$ can 
give potential higher-derivative terms in the effective action, 
when $p^2$'s act on the external background fields. Expanding $\wep$ 
around $({\Mvo})^{ij}$ gives the quadratic terms
\be\label{3.5}
\wep (\Mvo+\dM) = \wep(\Mvo) + \l~(t_2)^{ijk\ell}_{i'j'k'\ell'}
~(\Pf\Mvo)^{1/n} (\Mvo)^{-1}_{ij}(\Mvo)^{-1}_{k\ell} 
\dM^{i'j'}\dM^{k'\ell'}
+\cdots ,
\ee
where the numerical tensor $({t_2})^{ijk\ell}_{i'j'k'\ell'}$ controls 
how the $ij\ldots$ indices are contracted with the $i'j'\ldots$ indices. 
We will drop this tensor for now, though its form will be needed for a 
later argument.  For our immediate purposes it suffices to note that in 
the $\ve\to0$ limit the tensor structure 
of our tree diagrams is fixed by the $\Sp(2\nc)\times \SU(2\nf-2)$ 
subgroup of the global symmetry that is preserved by the vacuum.

Specializing to the massive modes, for which $\{i,j,k,\ell\}\to
\{u,v,w,x\}$, and using (\ref{deformedvac}) then gives the mass 
$m\sim \l\ve^{-\a}\m^\b$, where
\be\label{abeq}
\a:={n-1\over n}, \qquad\qquad
\b:={\nc\over n} .
\ee
The propagators are then
\bea\label{props}
\dM^{uv}\ \hbox{\bf--\, --\, --\, --}\ \dM^{wx} \quad &\sim &\quad 
{\ve^\a\over\l\m^\b} {D^2 \over p^2} 
\left(1+\left|\ve^\a\over\l\m^\b\right|^2 p^2\right)^{-1},
\nonumber\\
\dMb_{uv}\ \hbox{\bf-----------}\ \dMb_{wx} \quad &\sim &\quad 
{\vb^\a\over\lb\mb^\b} {\Db^2 \over p^2} 
\left(1+\left|\ve^\a\over\l\m^\b\right|^2 p^2\right)^{-1},
\nonumber\\
\dMb_{uv}\ \hbox{\bf-----\, --\, --}\ \dM^{wx} \quad &\sim &\quad 
\left|\ve^\a\over\l\m^\b\right|^2
\left(1+\left|\ve^\a\over\l\m^\b\right|^2 p^2\right)^{-1},
\eea
where have suppressed the tensor structures on the $\{u,v,w,x\}$ 
indices.

The (anti-)chiral vertices come from higher-order terms in the 
expansion of $\wep$ ($\bar\wep$).  Each (anti-)chiral vertex will 
have a $\Db^2$ ($D^2$) acting on all but one of its internal legs.  
Also, each vertex is accompanied by an $\int d^4\th$.  The 
$\ell$th-order term in the expansion of $\wep$ has the general 
structure 
\be\label{3.6}
\l~(t_\ell)^{i_1j_1\cdots i_\ell j_\ell}_{i_1'j_1'\cdots 
i_\ell' j_\ell'}~(\Pf\Mvo)^{1/n} 
(\Mvo)^{-1}_{i_1j_1}\cdots (\Mvo)^{-1}_{i_\ell j_\ell}
\dM^{i_1'j_1'}\cdots \dM^{i_\ell' j_\ell'},
\ee
where the numerical tensor $(t_\ell)^{i_1j_1\cdots i_\ell 
j_\ell}_{i_1'j_1'\cdots i_\ell' j_\ell'}$ controls how the 
$i_1j_1\cdots i_\ell j_\ell$ indices are contracted with the 
$i_1'j_1'\cdots i_\ell' j_\ell'$ indices.  Thus vertices with 
$m$ massless legs and $\ell-m$ massive legs are accompanied by 
the factors
\bea
\underbrace{\overbrace{\lower20pt
\hbox{\epsfig{file=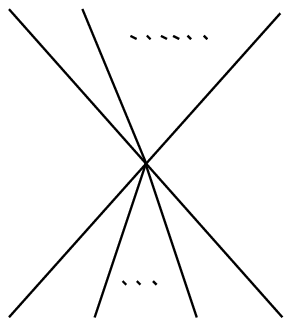,width=15mm}}}
^{m\ \mbox{\footnotesize{massless}}}}_{\ell-m\ 
\mbox{\footnotesize{massive}}}\quad \sim  		
\quad {\lb\over\vb^{\g_{\ell,m}}\mb^{\k_m}},\quad\qquad
\underbrace{\overbrace{\lower20pt
\hbox{\epsfig{file=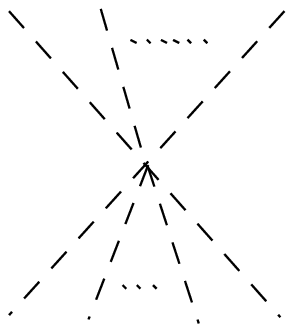,width=15mm}}}
^{m\ \mbox{\footnotesize{massless}}}}_{\ell-m\ 
\mbox{\footnotesize{massive}}}\qquad \sim  		
\quad {\l\over\ve^{\g_{\ell,m}}\m^{\k_m}},
\eea
where
\be\label{gkeq}
\g_{\ell,m} := \ell-{m\over2}-{n+1\over n},
\qquad\qquad
\k_m := {m\over2}-{\nc\over n}.
\ee
Note that it follows from (\ref{3.6}) that the number, $m$, of 
massless legs $\dM^{au}$ must be even.  This is because these legs 
each have one index $a\in\{1,2,\cdots 2\nc\}$ and the only 
non-vanishing components of $(\Mvo)^{-1}_{ij}$ with indices 
in this range are $(\Mvo)^{-1}_{ab} = -(\Mvo)^{-1}_{ba} = \m^{-1}$. 
Finally, to each (anti-)chiral external leg at zero momentum 
is assigned a factor of the (anti-)chiral background field 
$\dM^{au}(x,\th)$ ($\dMb^{au}(x,\thb)$) all at the same $x$. 
Overall momentum conservation means that the diagram has a 
factor of $\int d^4x$.  The $\d^4(\th-\th')$ for each internal 
propagator together with the $\int d^4\th$ integrals at each 
vertex leave just one overall $\int d^4\th$ for the diagram.

Before going on to the cases where the effective superpotentials 
are singular, we start by first looking at the $\nf=\nc+1$ and 
$\nf=\nc+2$ cases. These cases have regular superpotentials and 
are simple enough to show the details of the calculations. Although 
the superpotentials are regular in these cases we nevertheless 
expand them around the modified vacuum (\ref{deformedvac}), and 
then take the limit $\ve \to 0$ at the end. The purpose of doing 
the calculations around $\Mvo$ (rather than $M_0$) is to familiarize 
the reader with how the calculations will be implemented for singular 
superpotentials where expanding around the modified vacuum is 
necessary.

\subsection{$\bf \nf=\nc+1$} 

\FIGURE{\epsfig{file=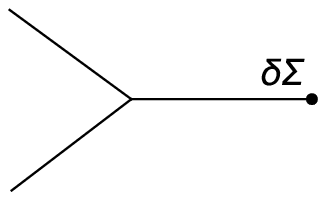,width=35mm}
\caption{The diagram reproducing the 
$F$-term for $\nf=\nc+1$ flavors.}}

This case is special since the superpotential is of a 
different form (\ref{2.2}), involving the Lagrange multiplier 
field $\Sigma$. Expanding (\ref{2.2}) around $\Mvo$, we have
\bea\label{3.5}
\wep &=& [\Pf\Mvo-\L^{2(\nc+1)}]~\dS 
+[(\Pf\Mvo)(\Mvo)^{-1}_{j i}]~\dM^{i j}\dS \nonumber\\  
&&\mbox{}+(\Pf\Mvo)\Bigl[ (\Mvo)^{-1}_{\ell k}(\Mvo)^{-1}_{j i}
+(\Mvo)^{-1}_{j \ell}(\Mvo)^{-1}_{k i}\nonumber\\
&&\qquad\qquad\qquad\qquad\mbox{} 
+(\Mvo)^{-1}_{j k}(\Mvo)^{-1}_{i\ell}\Bigr] \dM^{i j} \dM^{k \ell}\dS
+ \cdots
\nonumber\\
&=& [\Pf\Mvo-\L^{2(\nc+1)}]~\dS - \m^{\nc} \e_{uv}~\dM^{uv}\dS 
-\m^{\nc-1} \e_{u v} J_{a b}~\dM^{a u} \dM^{b v}\dS +\cdots ,
\nonumber
\eea

\noindent where we have just expressed the terms which are relevant in 
reproducing the multi-fermion $F$-term for $\nf=\nc+1$. Since 
the superpotential  includes the additional field $\Sigma$, we 
cannot use the coefficients for various superspace Feynman diagrams 
as expressed in (\ref{props}) and (\ref{gkeq}). Instead, reading 
the appropriate terms off the $\wep$ expansion, the propagator 
between $\dS$ and $\dM^{uv}$ is accompanied by a factor of 
$\e^{uv}/[p^2+ (\m\mb)^\nc]$, a vertex of $\dM^{au}\dM^{bv}\dS$ 
comes with a factor of $\e^{uv} J^{ab} \m^{\nc-1}$, and the $\dS$ 
vertex with a factor of $(\Pf\Mvo-\L^{2(\nc+1)})$. Evaluating the 
diagram in figure 3, we have
\bea\label{3.4n}
\d S &\sim& \int d^4x~d^4\th\, \left({{\Pf\Mvo-\L^{2(\nc+1)}}\over {\mb\m^{\nc-1}}}\right)\e^{uv}J^{ab}
\,\dMb_{au} \cdot\dMb_{bv}. 
\eea
In the $\ve\to 0$ limit, $\Pf\Mvo$ vanishes leaving us with 
\bea\label{3.4n}
\d S &\sim& \int d^4x~d^2\th\, {1\over {\mb\m^{\nc-1}}}~\e^{uv}J^{ab}
\,(\Db\dMb_{au} \cdot \Db\dMb_{bv}), 
\eea
where we have traded a $\int d^2\thb$ for a $\Db^2$ and used the 
equation of motion $\Db^2\dMb=0$ to leading order in $\dMb$ to distribute 
the $\Db$'s amongst $\dMb$'s.  The above expression for 
$\d S$ is the higher-derivative $F$-term for $\nf=\nc+1$. 

\subsection{$\bf \nf=\nc+2$} 

\FIGURE{\epsfig{file=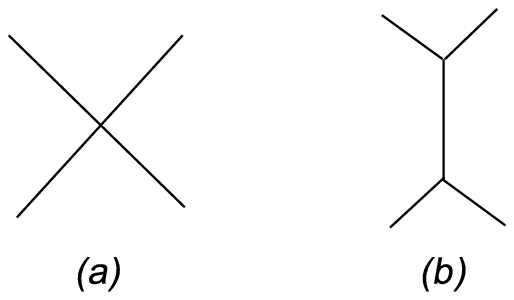,width=45mm}
\caption{Diagrams with four massless 
external anti-chiral legs for $\nf=\nc+2$. (a) The amputated 
4-vertex which does not have the right structure. (b) This diagram reproduces the multi-fermion 
F-term.}}

In order to reproduce the $F$-term for  $\nf=\nc+2$ 
flavors we need four massless anti-chiral legs.  There are only 
two such diagrams, shown in figure 4.  Diagram (a) 
with an amputated 4-vertex ($m=l=4$) does not have the right 
structure to be an $F$-term because, in the $\ve\to 0$ 
limit, it contributes to the action the term
\bea\label{n1firstdiag}
&& \int d^4x~d^4\th\, {\lb\over {\mb^{2-\nc}}}~ 
A^{abcd}_{a'b'c'd'}~\e^{uvwx} J^{a'b'}
J^{c'd'}\nonumber\\
&&\qquad\qquad\qquad\mbox{}\times
\dMb_{au}\dMb_{bv} \dMb_{cw} \dMb_{dx},
\eea
where $A^{abcd}_{a'b'c'd'}$ is a non-vanishing tensor which 
determines how $ab\cdots$ indices are contracted with $a'b'\cdots$ 
indices.  Even if we traded a $\int d^2\thb$ for a $\Db^2$ and 
distributed the $\Db$'s among $\dMb_{au}$'s, we would still need 
another $\Db^2$.  Also, the coefficient in the 
integrand of (\ref{n1firstdiag}) does not match that of 
(\ref{mfftlead}) for $\nf=\nc+2$.   This term is probably
just a correction to the K\"ahler potential, though we have
not ruled out the possibility that it is a new global 
$F$-term different from (\ref{bwgen}).  

Diagram (b) consists of two external 
anti-chiral vertices and one anti-chiral propagator, and
gives 
\bea\label{n1seconddiag}
\d S &\sim& \int d^4x~d^4\th_1d^4\th_2\ 
\dMb_{au}(\th_1)\dMb_{bv}(\th_1) J^{ab}(J^{us}J^{vt}-J^{ut}J^{vs})
{\lb\over\vb^{\g_{3,2}}\mb^{\k_2}}\nonumber\\
&&\ \ \mbox{}\times \d^4(\th_1-\th_2) (J_{sp}J_{tq}-J_{sq}J_{tp})
{\vb^\a\over\lb\mb^\b}{\Db^2\over p^2} 
\left(1+\left|\ve^\a\over\l\m^\b\right|^2p^2\right)^{-1}\nonumber\\
&&\ \ \mbox{}\times {\lb\over\vb^{\g_{3,2}}\mb^{\k_2}}
(J^{wp}J^{xq}-J^{wq}J^{xp})J^{cd}~\dMb_{cw}(\th_2)\dMb_{dx}(\th_2).
\eea
Using the values $\a=0$, $\b=\nc$, $\g_{3,2}=0$ and $\k_2=1-\nc$, and 
substituting them in (\ref{n1seconddiag}), we obtain
\bea\label{n1seconddiagcont}
\d S &\sim & \int d^4x~d^4\th \, \dMb_{au}\dMb_{bv} {\lb\over \mb^{2-\nc}} 
{\Db^2\over p^2} \left[1-|\l \m^{\nc}|^{-2}p^2 + {\cal O}(p^4)\right] 
\dMb_{cw} \dMb_{dx}
\nonumber\\ 
&=& {\lb\over\mb^{2-\nc}}~\e^{uvwx}J^{ab}J^{cd}\int d^4x~d^2\thb\ 
\dMb_{au}\dMb_{bv} {D^2\Db^2\over p^2}(\dMb_{cw}\dMb_{dx})\nonumber\\
&&\mbox{}-{\e^{uvwx}J^{ab}J^{cd}\over\l\m^{\nc}\mb^2}\int d^4x~d^2\th\ 
\Db^2\left[  \dMb_{au}\dMb_{bv}\Db^2(\dMb_{cw}\dMb_{dx})\right]
+{\cal O}(p^2) 
\nonumber\\
&=& {\lb\over\mb^{2-\nc}}~\e^{uvwx}\int d^4x~d^2\thb\ 
\dMb_{au}\dMb_{bv}\dMb_{cw}\dMb_{dx} 
\nonumber\\
&&\mbox{}-{\e^{uvwx}\over\l\m^{\nc}\mb^2}\int d^4x~d^2\th\ 
(\Db\dMb_{au}\cdot\Db\dMb_{bv})(\Db\dMb_{cw}\cdot\Db\dMb_{dx})
+{\cal O}(p^2), 
\eea
where in the second line in (\ref{n1seconddiagcont}) we have traded 
an $\int d^2\th$ for a $D^2$  and used the identity $D^2\Db^2=p^2$ 
on antichiral fields to cancel the IR pole. We then traded an $\int 
d^2\thb$ for a $\Db^2$ in the third line and used the equation of 
motion $\Db^2\dMb=\cal O (\dMb)$ to distribute the $\Db$'s in the 
fifth line. In the last two lines, the first term in the expansion 
does not have the right structure the multi-fermion $F$-term, but 
the second term is, up to some numerical factor, 
the multi-fermion $F$-term in (\ref{mfftlead}) for $\nf=\nc+2$. 
All other diagrams in the expansion vanish in the limit $p\to 0$.

\subsection{$\bf \nf=\nc+3$}

This is the first case where we have a singular superpotential. 
The $F$-term for $\nf=\nc+3$ has six external anti-chiral massless 
legs so we have to look for those Feynman diagrams with only six external 
anti-chiral legs. There are five different possibilities (plus 
their crossed-channels), see figure 5. Among these diagrams, The 
four diagrams in figure 5 (a) either do not have the right structure 
to be a multi-fermion $F$-term, or have zero coefficient.  For example, 
the second graph from the left in figure 5 (a) comes with zero coefficient 
because it has vertices with an odd number of massless legs.  The rest 
of diagrams in figure 5 (a) are probably corrections to the K\"ahler term,
though we have not ruled out the possibility that some of them might
contribute to new classes of global $F$-terms different from (\ref{bwgen}).

\FIGURE{\epsfig{file=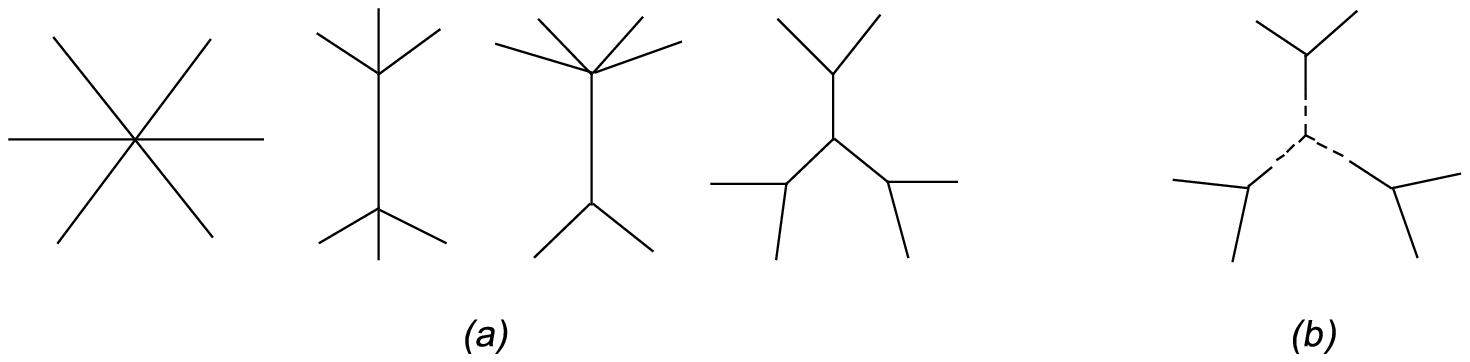,width=125mm}
\caption{Diagrams with six massless external anti-chiral legs for 
$\nf=\nc+3$. (a) Diagrams which do not have the right structure. 
(b) The only diagram contributing to the $F$-term (\ref{mfftlead}).}}

The only diagram with the right structure is 5 (b): three external 
anti-chiral vertices with one chiral internal vertex.  Evaluating 
this diagram gives
\bea\label{n2mixeddiag}
\d S &\sim & \int\!\! d^4x\, d^4\th B^{uvwxyz}_{u'v'w'x'y'z'}\,
J^{u'v'}J^{w'x'} J^{y'z'} J^{ab}\,
\dMb_{au}\dMb_{bv} {\lb\over\vb^{\g_{3,2}}\mb^{\k_2}}
\left|\ve^\a\over\l\m^\b\right|^2
\left(1+\left|\ve^\a\over\l\m^\b\right|^2 p^2\right)^{-1}\nonumber\\
&&\qquad\qquad\mbox{}\times {\l\over\ve^{\g_{3,0}}\m^{\k_0}}
\left|\ve^\a\over\l\m^\b\right|^2
\left(1+\left|\ve^\a\over\l\m^\b\right|^2 p^2\right)^{-1} 
{\lb\over\vb^{\g_{3,2}}\mb^{\k_2}}
J^{cd}~\Db^2 (\dMb_{cw}\dMb_{dx})\nonumber\\
&&\qquad\qquad\mbox{}\times \left|\ve^\a\over\l\m^\b\right|^2
\left(1+\left|\ve^\a\over\l\m^\b\right|^2 p^2\right)^{-1} 
{\lb\over\vb^{\g_{3,2}}\mb^{\k_2}}
J^{ef}\,\Db^2 (\dMb_{ey}\dMb_{fz}),
\eea
with $B^{uvwxyz}_{u'v'w'x'y'z'}$ being a tensor contracting $uv\cdots$ 
indices to $u'v'\cdots$ indices.   Substituting the values $\a={1\over2}$, 
$\b={\nc\over2}$, $\k_0=-{\nc\over2}$, $\k_2=1-{\nc\over2}$, $\g_{3,0}= 
{3\over2}$ and $\g_{3,2}= {1\over2}$ into (\ref{n2mixeddiag}) and taking 
the limit $\ve\to 0$, we obtain
\bea\label{n2mixeddiagcont}
\d S \sim \int d^4x~d^2\th &&{\e^{uvwxyz}\over{\l^2\mb^3 \m^{\nc}}}~ 
J^{ab} J^{cd} J^{ef} (\Db\dMb_{au}\cdot\Db\dMb_{bv})\nonumber\\
&&(\Db\dMb_{cw}\cdot\Db\dMb_{dx})(\Db\dMb_{ey}\cdot\Db\dMb_{fz}),
\eea
where we have used the fact that in the $\ve\to 0$ limit the flavor 
symmetry group is enhanced to $\Sp(2\nc) \times \SU(2\nf-2\nc)$.  This 
expression coincides with (\ref{mfftlead}) for $\nf=\nc+3$. 
Since this was the only diagram contributing in the $\nf=\nc+3$ case, 
there can be no cancellation of its coefficient.  This shows that the 
$\nf=\nc+3$ singular superpotential indeed reproduces the corresponding 
higher-derivative global F-term in perturbation theory.  

\subsection{$\bf \nf\geq\nc+4$}

As we go higher in the number of flavors, however, the number of 
diagrams contributing to each amplitude rapidly increases.  For instance, 
just among the class of internally purely-chiral diagrams illustrated 
in figure 6, there are are four superspace Feynman diagrams in the 
case of $\nf=\nc+4$ flavors each with the right structure to contribute
to (\ref{mfftlead}). 
But since now multiple diagrams contribute, we must show in addition that no 
cancellations occur that could set the coefficient of the higher-derivative 
term to zero.  This seems quite complicated, as it depends on the signs 
and tensor structures of the vertices.  Some sort of symmetry argument is 
clearly wanted, but still eludes us.

\FIGURE{\epsfig{file=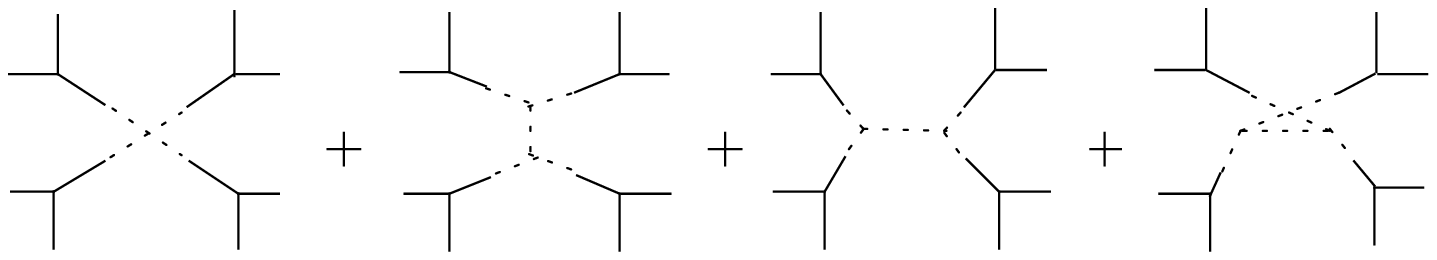,width=130mm}
\caption{Diagrams which have the right structure to give a 
higher-derivative $F$-term for $\nf=\nc+4$.}}

In addition, there are now also other classes of diagrams which are 
neither purely anti-chiral (as in figure 5(a)) or internally purely 
chiral (as in figure 6). 
It is not clear whether these mixed diagrams will also contribute to 
higher-derivative amplitudes of the form (\ref{mfftlead}) or not.

\section*{Acknowledgments}
It is a pleasure to thank C. Beasely, S. Hellerman, N. Seiberg, and 
E. Witten for helpful comments and discussions. This work is supported 
in part by DOE grant DOE-FG02-84ER-40153.

\end{document}